\documentclass[reprint,superscriptaddress,nofootinbib,amsmath,amssymb,prd,floatfix]{revtex4}
\pdfoutput=1
\usepackage{graphicx}
\usepackage[dvipsnames]{xcolor}
\usepackage{amsmath}
\allowdisplaybreaks
\usepackage{amssymb}

\usepackage{slashed}
\usepackage[utf8]{inputenc}
\usepackage[colorlinks=true,linkcolor=blue,bookmarksopen,bookmarksnumbered]{hyperref}
\usepackage{color}
\usepackage[normalem]{ulem}
\usepackage{soul}
\usepackage{units}
\usepackage{rotating}
\usepackage{hhline,multirow,tabularx}
\usepackage{bm}
\usepackage{hyperref}
\usepackage[export]{adjustbox}
\usepackage{mdframed}
\usepackage{comment}
\usepackage{natbib}

\def\bea#1\eea{\begin{align}#1\end{align}}

\usepackage{lipsum}

\begin{document}

\title{Lepton pair production in UPCs: towards the precision test of the  resummation 
formalism}

\author{Ding Yu Shao}
\affiliation{Department of Physics, Center for Field Theory and Particle Physics, Key Laboratory of
Nuclear Physics and Ion-beam Application (MOE),  Fudan University, Shanghai, 200433, China}
\affiliation{Shanghai Research Center for Theoretical Nuclear Physics, NSFC and Fudan University, Shanghai 200438, China}

\author{Cheng Zhang}
\affiliation{Department of Physics, Center for Field Theory and Particle Physics, Key Laboratory of
Nuclear Physics and Ion-beam Application (MOE),  Fudan University, Shanghai, 200433, China}

\author{Jian~Zhou}
 \affiliation{\normalsize\it Key Laboratory of
Particle Physics and Particle Irradiation (MOE),Institute of
Frontier and Interdisciplinary Science, Shandong University,
QingDao, China }

\author{Ya-jin Zhou}
\affiliation{\normalsize\it Key Laboratory of Particle Physics and
Particle Irradiation (MOE),Institute of Frontier and
Interdisciplinary Science, Shandong University, QingDao,  China }


\begin{abstract}

We present a detailed investigation of the azimuthal asymmetries and acoplanarity in lepton pair production in ultraperipheral collisions (UPCs). These observables provide a unique opportunity to test the SCET resummation formalism, given the extremely high photon flux in UPCs, which enables precise measurements of these processes. We improve the accuracy of the previous calculations by including the soft photon contributions beyond the double leading logarithm approximation. Notably, the single logarithm terms arising from the collinear region are greatly enhanced by the small mass of the leptons. Our findings demonstrate the accessibility of these sub-leading resummation effects through the analysis of angular correlations in lepton pairs produced in UPCs at the Relativistic Heavy Ion Collider (RHIC) and the Large Hadron Collider (LHC).

\end{abstract}
\maketitle

\section{Introduction}
The  study of pure electromagnetic di-lepton production in  ultraperipheral heavy ion collisions (UPCs) has been and continues to be an active area of research  since the physics operation began at Relativistic Heavy Ion Collider (RHIC)~\cite{Baltz:2007kq,Bertulani:2005ru,STAR:2018ldd,STAR:2018xaj,ATLAS:2018pfw,ALICE:2018ael,Klein:2016yzr,ALICE:2022bii,Zhou:2022gbh,Wang:2022ihj,Niu:2022cug}.   One of the key features of di-lepton production in UPCs is the enhancement of the cross-section by a factor of $Z^4$ at low transverse momentum, where $Z$ is the nuclear charge number.  This enhancement makes di-lepton production in UPCs an attractive channel for searching for physics beyond the Standard Model~\cite{DELPHI:2003nah,ATLAS:2022ryk,Knapen:2016moh,CMS:2018erd,Ellis:2017edi,Xu:2022qme}, as well as for studying the properties of QED under extreme conditions~\cite{Baur:1998ay,Klein:2020fmr,Steinberg:2021lfm,Hattori:2020htm,Copinger:2020nyx,Brandenburg:2021lnj}.  In addition, di-lepton production serves as a baseline measurement for the EM probes of the QGP, which are essential for understanding its properties.

Di-lepton back-to-back production in UPCs recently gained the renewed interest from both experimental and theoretical sides. This is  partially triggered by the observation of the lepton pair transverses momentum $q_\perp$ broadening at LHC~\cite{ATLAS:2018pfw,ALICE:2018ael,ALICE:2022bii} and RHIC~\cite{STAR:2018ldd,STAR:2018xaj}.  The mean value of lepton pair transverse momentum   was found to increase with decreasing impact parameter $b_\perp$ which is the transverse distance between two colliding nucleus.  To account for this phenomenon, it is crucial to employ a formalism~\cite{Vidovic:1992ik,Hencken:1994my,Zha:2018tlq,Klein:2020jom,Wang:2021kxm,Wang:2022gkd,Lin:2022flv,Wang:2022ihj,Klusek-Gawenda:2020eja} that allows us to derive the joint $b_\perp$ and $q_\perp$ dependent cross section. As a result, the coherent photon distribution enters the cross section formula is the Wigner distribution rather than the transverse momentum dependent (TMD) distribution.

In this paper, we focus on  relatively high pair transverse momentum region where $q_\perp$ spectrum is no longer controlled by the primordial coherent photon distribution.
 The $q_\perp$ spectrum instead is dominantly  generated via the recoiled effect due to the final state soft photon radiations when $q_\perp$ is much larger than the reverse of nuclear radius. Despite the smallness of the fine coupling constant $\alpha_e$, the fixed order contribution is greatly enhanced by the large logarithm term of the type $\frac{\alpha_e}{\pi} {\rm ln} \frac{M^2}{m^2}  {\rm ln}\frac{P_\perp^2}{q_\perp^2}$ and thus call for a resummation, where $M$ and $m$ are the invariant mass of the lepton pair and lepton mass respectively, and $P_\perp$ is approximately the individual lepton transverse momentum. 
Such a resummation formalism was first developed  in the context of  heavy quark pair production~\cite{Zhu:2012ts,Li:2013mia,Catani:2014qha,Ju:2022wia}. It later has been extended to include azimuthal dependent contributions~\cite{Hatta:2020bgy,Hatta:2021jcd} and applied to study azimuthal asymmetries in di-lepton production in UPCs~\cite{Hatta:2021jcd,Shao:2022stc} in the  leading double logarithm approximation.  The first attempt to take into account the sub-leading logarithm contribution relevant in the kinematic region where  $m$ is of the same order of $M$ to the azimuthal asymmetries in di-muon production has been presented in Ref.~\cite{Shao:2022stc}.  The next to leading logarithm contribution to azimuthal asymmetries turns out to be sizable for muon production case.
The purpose of this work is to resum the sub-leading logarithm contribution  in di-electron production which is important in the kinemaitc limit $m \ll M$. Note that  at low $q_\perp$, the azimuthal angular  correlation in di-lepton pair production  mainly arises from the linear polarization of coherent photons~\cite{Li:2019yzy,Li:2019sin,Adam:2019mby,Xiao:2020ddm,Zhao:2022dac,Brandenburg:2022tna} rather than  soft photon radiation effect.

 Besides directly measuring $q_\perp$ distribution, the azimuthal
angular decorrelation of the lepton pair is often experimentally studied as well.  When the lepton pair acquires finite transverse momentum  either from the incoming coherent photons or from the recoiled effect due to soft photon radiation, electron and positron  are no longer  produced in the exact back-to-back configuration in the transverse momentum phase space. The degree of the deviation from the back-to-back configuration is measured by the quantity so called acoplanarity whose definition will be specified later. The acoplanarity  as the way of exploring  nucleon  structure and QGP properties was extensively discussed  in the context of the di-jet production and gauge boson-jet production \cite{Banfi:2008qs,Hautmann:2008vd,Zheng:2014vka,Sun:2014gfa,Sun:2015doa,Chen:2018fqu,Sun:2018icb,Liu:2018trl,Chien:2019gyf,Liu:2020dct,Chien:2020hzh,Abdulhamid:2021xtt,Chien:2022wiq,Bouaziz:2022tik,Yang:2022qgk,Martinez:2022dux,Zhang:2020onw}. The acoplanarity in di-lepton production in UPCs was first studied  in Refs.~\cite{Klein:2018fmp,Klein:2020jom}.  The leading double logarithm involved in the calculation of  this observabl is the type of $\frac{\alpha_e}{2\pi}  \ln^2 \frac{M^2}{q_x^2}$, where $q_x$ is one of the transverse components of $q_\perp$ perpendicular to $P_\perp$. In this work, we extend the resummation formalism to the next to leading logarithm accuracy and investigate its phenomenological consequence as well.

The paper is structured as follows. We first briefly review the previous calculations for the observables under consideration in the next section. The resummation formalism  formulated in the effective theory  are discussed  in Sec.III and Sec.IV for two different kind of angular correlations.  The approach based on the effective theory allows us to resum the sub-leading logarithm contribution to all orders in a systematic manner.  We also present the heuristic derivations of the two Sudakov factors up to the next to leading logarithm accuracy following a more conventional perturbative QCD method in the appendices.  The numerical results are presented in section V. We summarize the paper in section VI.

\section{Angular correlations in the leading double logarithm approximation }\label{sec:DL}
At low total transverse momentum of di-lepton pair, 
 electron and positron pair is dominantly produced
via two coherent photon fusion process. The corresponding kinematics are specified as the follows.
\begin{eqnarray}
\gamma_1(x_1P+k_{1\perp})+\gamma_2(x_2 \bar P+k_{2\perp}) \rightarrow l^+(p_1)+ l^-(p_2).
\end{eqnarray}
The leptons are produced nearly back-to-back in azimuthal with  total transverse momentum
$q_\perp\equiv p_{1\perp}+p_{2\perp}=k_{1\perp}+k_{2\perp}$
being much smaller than  $P_\perp=(p_{1\perp}-p_{2\perp})/2$.   To sort out the UPC events, one has to first compute the impact parameter dependent cross section~\cite{Vidovic:1992ik,Hencken:1993cf} and then integrate  $b_\perp$
 over the range $[2R_{WS}, \infty)$, where $b_\perp$ is the transverse distance between two colliding nuclei and $R_{WS}$ is the nuclear radius.  Once $b_\perp$ dependence is introduced, the transverse momentum carried by the incoming photon in the amplitude is no longer identical to that in the conjugate amplitude.  Below we use $k_{1\perp}$, $k_{2\perp}$ and $k_{1\perp}'$, $k_{2\perp}'$ to denote transverse momenta in the amplitude  and transverse momenta in the conjugate amplitude with the constraint
 $ k_{1\perp}'+k_{2\perp}'\equiv q_\perp$.
The Born cross section of the di-electron production takes the form~\cite{Li:2019yzy,Li:2019sin,Pisano:2013cya},
 \begin{eqnarray}
\frac{d\sigma_{\!_0}}{d^2 p_{1\perp} d^2 p_{2\perp} dy_1 dy_2 d^2 b_\perp }= \frac{2\alpha_e^2}{M^4}\frac{1}{(2\pi)^2}
\left [ \mathcal{A}+{\mathcal B} \cos 2\phi+\mathcal{C} \cos 4\phi \right ],
\end{eqnarray}
where $\phi$ is the angle between transverse momenta $q_\perp$ and
$P_\perp$. $y_1$ and $y_2$ are leptons rapidities, respectively.  M is the invariant mass of the lepton pair.  At low $q_\perp$, the  $\cos 4\phi$ azimuthal modulation  is mainly induced by the linear polarization of coherent photons~\cite{Li:2019yzy,Li:2019sin,Adam:2019mby,Xiao:2020ddm,Zhao:2022dac,Brandenburg:2022tna}. The computed  $\cos 4\phi$ asymmetry~\cite{Li:2019yzy,Li:2019sin}   is in excellent agreement with the measured asymmetries by STAR collaboration~\cite{Adam:2019mby}. It is worthy to mention that the polarization dependent reactions in UPCs opens a new avenue to explore the novel QCD phenomenology
~\cite{Hagiwara:2020juc,Xing:2020hwh,Hagiwara:2021qev,Brandenburg:2022jgr,Mantysaari:2022sux,Wu:2022exl,STAR:2022wfe,Zha:2020cst}. The hard coefficient $\mathcal B$ is suppressed by the power of $\frac{m^2}{M^2}$ at the tree level, and is neglected. 
 The coefficients $\mathcal{A}$ and $\mathcal{C}$ have been computed at the leading order in Ref.~\cite{Li:2019sin},
 \begin{eqnarray}
\mathcal{A}&=& \frac{M^2-2 P_\perp^2}{P_\perp^2}\frac{Z^4
\alpha_e^2}{\pi^4}\int d^2k_{1\perp} d^2 k_{2\perp} d^2
\Delta_\perp \delta^2( q_\perp-k_{1\perp}-k_{2\perp}) e^{i
\Delta_\perp \cdot b_\perp}
\nonumber \\&& \ \ \ \ \ \ \ \ \ \ \ \ \ \ \ \ \ \ \times \left [
(k_{1\perp} \cdot k_{1\perp}')(k_{2\perp} \cdot k_{2\perp}')+
(k_{1\perp} \! \cdot k_{2 \perp})\Delta_\perp^2
-(k_{1\perp}\!\cdot \Delta_{ \perp})(k_{2\perp}\!\cdot \Delta_{ \perp})
\right ]
\nonumber \\&& \ \ \ \ \ \ \ \ \ \ \ \ \ \ \ \ \ \ \times \
{\cal F}(x_1,k_{1\perp}^2){\cal F}^*(x_1,k_{1\perp}'^2){\cal F}(x_2,k_{2\perp}^2){\cal F}^*(x_2,k_{2\perp}'^2),
\end{eqnarray}
and
 \begin{eqnarray}
\mathcal{C} =
 -2 \frac{Z^4\alpha_e^2}{\pi^4}
\!\!\!&&\!\! \!\!\!\!  \int d^2k_{1\perp} d^2 k_{2\perp} d^2\Delta_\perp
\delta^2( q_\perp-k_{1\perp}-k_{2\perp})  e^{i\Delta_\perp \cdot b_\perp}
\nonumber \\
&\times& \left \{2\left [ 2( k_{2\perp} \! \cdot \hat q_{\perp})( k_{1\perp} \! \cdot \hat q_{\perp})
- k_{1\perp} \!\cdot \! k_{2\perp} \! \right ]
\left [ 2( k_{2\perp}' \! \cdot \hat q_{\perp})( k_{1\perp}' \! \cdot \hat q_{\perp})
- k_{1\perp}' \!\cdot \! k_{2\perp}' \! \right ] \right .\
\nonumber \\ && \left .\ -\left [
(k_{1\perp} \cdot k_{1\perp}')(k_{2\perp} \cdot k_{2\perp}')+
(k_{1\perp} \! \cdot k_{2 \perp})\Delta_\perp^2
-(k_{1\perp}\!\cdot \Delta_{ \perp})(k_{2\perp}\!\cdot \Delta_{ \perp})
\right ]\right \}
\nonumber \\
& \times &
{\cal F}(x_1,k_{1\perp}^2){\cal F}^*(x_1,k_{1\perp}'^2){\cal F}(x_2,k_{2\perp}^2){\cal F}^*(x_2,k_{2\perp}'^2),
\end{eqnarray}
where $\Delta_\perp=k_{1\perp}-k_{1\perp}'=k_{2\perp}'-k_{2\perp}$.
 $\hat q_\perp$ is unit vector defined as  $\hat q_\perp= q_\perp/| q_\perp| $.
The incoming photons  longitudinal momenta fraction  are fixed by the
external kinematics according to   $x_1=\sqrt{\frac{P_\perp^2+m^2}{s}}(e^{y_1}+e^{y_2})$
 and $x_2=\sqrt{\frac{P_\perp^2+m^2}{s}}(e^{-y_1}+e^{-y_2})$ with $m$ being the lepton mass and $s$ being
 the center mass energy.  ${\cal F}(x,k_\perp^2)$ describes the amplitude of finding a photon carrying the certain momentum. For a given nuclear charge form factor $F(k^2)$, it is computed as
${\cal F}(x,k_\perp^2)=\frac{F(k_{\perp}^2+x^2M_p^2)}{(k_{\perp}^2+x^2M_p^2)}$,
 where $M_p$ is proton mass. Note that the lepton mass is ignored in the hard coefficients.

At low $q_\perp$, the pair transverse momentum is mainly determined by the primordial coherent photon distributions, while at high $q_\perp$, the transverse momentum spectrum is dominantly produced perturbatively  via final state soft photon radiation effect. Such soft photon radiation contribution is enhanced by the large logarithms and need to be resummed to all orders. It is the most convenient to achieve the all order resummation in the transverse position space,
 \begin{eqnarray}
  \frac{d\sigma}{d^2 p_{1\perp} d^2 p_{2\perp} dy_1 dy_2 d^2 b_\perp }= \int
  \frac{d^2 r_\perp}{(2\pi)^2} e^{i r_\perp \cdot q_\perp} e^{- \mathrm{Sud}(r_\perp)} \int d^2 q_\perp'
  e^{-i r_\perp \cdot q_\perp'} \frac{d\sigma_{_{\!0}}(q_\perp')}{ d\mathcal{P.S.}}, \label{res1}
  \end{eqnarray}
  where $ \mathrm{Sud}(r_\perp)$ is normally referred to as the Sudakov factor, and the phase space $d\mathcal{P.S.}=d^2 p_{1\perp} d^2 p_{2\perp} dy_1 dy_2 d^2 b_\perp$ .  The leading logarithm contribution of the Sudakov factor has been derived in Ref.~\cite{Zhu:2012ts,Li:2013mia,Catani:2014qha,Hatta:2020bgy,Hatta:2021jcd},
  \begin{eqnarray}
 \mathrm{Sud}(r_\perp)=
\frac{\alpha_e}{\pi} {\rm ln} \frac{M^2}{m^2}  {\rm ln}\frac{P_\perp^2}{\mu_r^2},
\label{eq:sud_DL}
  \end{eqnarray}
  with $\mu_r=2 e^{-\gamma_E}/|r_\perp|$. In this work, we extended the all order Sudakov resummation to include the sub-leading logarithm contributions. The detailed derivations are presented in the next section (for a heuristic derivation, see Appendix A).

We now turn to discuss how to  formulate the calculation of the acoplanarity.  One can define an azimuthal angle $\phi_\perp =\pi -( \phi_1 - \phi_2)$ where $\phi_1$ and $\phi_2$ represent the azimuthal
angles for the lepton and the anti-lepton, respectively. 
The acoplanarity observed in experiments is defined as $\alpha= |\phi_\perp|/\pi $. We  fix the direction of electron transverse momentum $p_{1\perp} $ to be the Y-axis.  The acoplanarity can then be easily reconstructed by  the ratio of  $q_x$ (the  component of $q_\perp$ aligned with X-axis) and $P_\perp$.  The $q_x$ dependent cross section takes the form,
\begin{eqnarray}
  \frac{d\sigma}{dq_x d^2 P_{\perp} dy_1 dy_2 d^2 b_\perp }&=& \int dq_y
  \frac{dr_y d r_x}{(2\pi)^2} e^{i( r_x  q_x+r_y  q_y)} e^{-  \mathrm{Sud_a}(r_x,r_y)} \int d q_x' dq_y' \ 
  e^{-i (r_x  q_x'+r_y  q_y')} \frac{d\sigma_{_{\!0}}(q_\perp')}{ d\mathcal{P.S.}} \nonumber \\
  &=& \int 
  \frac{d r_x}{2\pi} e^{i r_x  q_x} e^{-  \mathrm{Sud_a}(r_x,r_y=0)} \int d q_x' dq_y' \ 
  e^{-i r_x  q_x'} \frac{d\sigma_{_{\!0}}(q_\perp')}{ d\mathcal{P.S.}}.
  \end{eqnarray}
  where the leading logarithm contribution to the Sudakov factor $ \mathrm{Sud_a}(r_x)$ was given by,
\begin{align}\label{eq:suda_DL}
 \mathrm{Sud_a}(r_x) = \frac{\alpha_e}{2\pi} \left[ \ln^2 \frac{M^2}{\mu_{rx}^2}  - \ln^2 \frac{m^2}{\mu_{rx}^2}  \theta(m-\mu_{rx})\right] ,
\end{align}
 with $\mu_{rx}=2 e^{-\gamma_E}/|r_x|$. This expression is identical to what is obtained in Ref~\cite{Klein:2018fmp} if $\mu_{rx}$ is replaced with $\mu_r$. One notices that we make a one-dimensional Fourier transform in the above resummed formula rather than a two-dimensional Fourier transform as has been done in Eq.~\eqref{res1}. This is because the acoplanarity is essentially a one-dimensional observable, whereas $q_\perp$ spectrum is a two-dimensional distribution. Naturally, the associated Sudakov factors in the two resummed cross sections differ from each other.    When deriving the momentum space expression of the Sudakov factor $ \mathrm{Sud_a}(l_x)$, the Y-component of soft photon transverse momentum has to be integrated over the whole available phase space region. It is thus not appropriate to reconstruct the $q_x$ distribution from the resummed $q_\perp$ distribution by integrating out $q_y$. We will present the detailed derivation of the Sudakov factor $ \mathrm{Sud_a} (r_x)$ in Section IV (for a heuristic derivation, see Appendix B).

\section{Mass factorization and resummation in SCET}

In the previous section, we presented the double logarithmic resummation formula for low transverse momentum $q_\perp$ and acoplanarity $\alpha$. In this section, we will utilize the SCET \cite{Bauer:2000yr,Bauer:2001ct,Bauer:2001yt,Bauer:2002nz,Beneke:2002ph} and standard RG methods to derive a resummation formula that includes the effects of lepton mass resummation to all orders. We adopt dimensional regularization in $d=4-2\epsilon$ dimensions, following the $\overline{\rm MS}$ prescription. The dimensional regularization scale $\mu^2$ is replaced by $\mu^2 e^{\gamma_E}/(4\pi)$, and we subtract $\epsilon$-poles to obtain the renormalized results in the $\overline{\rm MS}$ scheme. All renormalized results are presented in the $\overline{\rm MS}$ scheme.

\subsection{Factorization formula at low $q_\perp$}

In Ref. \cite{Shao:2022stc}, we derived a resummation formula at low transverse momentum $q_\perp$ for muon pair production at the RHIC energy by assuming $M\sim m \gg q_\perp$. The differential cross section is factorized into the product of hard and soft factors, with the lepton mass $m$ retained in both factors.

To obtain a resummation formula that includes lepton mass resummation, we need to refactorize the massive hard and soft functions in the small mass limit ($M \gg q_\perp \gtrsim m$). Explicitly, the massive hard function $H(M,m,\mu)$ is factorized as the product of the massless hard function $H(M,\mu)$ and collinear jet functions $J(m,\mu)$. Similarly, the massive soft function $S(l_\perp,\Delta y,m,\mu)$ is factorized as the product of the massless soft function $S(l_\perp,\Delta y,\mu)$ and collinear-soft functions $C_{i}(k_{i,\perp},p_T,m,\mu)$. The resulting differential cross section is given by
\begin{align}\label{eq:fac_qT}
	\frac{d\sigma(q_\perp)}{ d\mathcal{P.S.}} = & \, H(M,\mu) J^2(m,\mu) \int d^2 l_\perp d^2 k_{1\perp} d^2 k_{2\perp}  \frac{d\sigma_0(q_\perp-l_\perp - k_{1\perp }- k_{2\perp} )}{ d\mathcal{P.S.}}  \notag \\
 &\times S(l_\perp,\Delta y,\mu) C_1 ( k_{1\perp },P_\perp,y_1,m,\mu ) C_2 ( k_{2\perp },P_\perp,y_2,m,\mu ),  
\end{align}
where the hard function $H(M,\mu)$ comes from the matching from QED to the low energy effective theory, and it can be obtained from the virtual corrections for massless amplitudes of $\gamma\gamma\to l^+ l^-$. The corresponding anomalous dimension is written as
\begin{align}
    \Gamma_H &= \frac{\alpha_e}{4\pi} \left( 8 \ln \frac{M^2}{\mu^2} - 12\right), 
\end{align}
where the scale-dependent term gives double logarithmic resummation results, while the scale-independent term controls single logarithmic resummation. The physical scale in the hard function is $\mu_h=M$.

The collinear jet functions $J(m,\mu)$, which depend on the lepton mass, have been extensively studied in literatures \cite{Neubert:2007je,Ferroglia:2013awa,Bauer:2013bza,Fickinger:2016rfd,Kang:2020xgk,Gaggero:2022hmv,vonKuk:2023jfd}. In particular, the two-loop expression for these functions was derived in \cite{Mitov:2006xs,Becher:2007cu,Jain:2008gb}. At the one-loop level, the jet function takes the form
\begin{align}
	J^{\mathrm{NLO}}(m,\mu)=1+\frac{\alpha_e}{4 \pi}\left[\frac{2}{\epsilon^2}+\frac{1}{\epsilon}\left(1+2 \ln \frac{\mu^2}{m^2}\right)+\left(1+\ln \frac{\mu^2}{m^2}\right) \ln \frac{\mu^2}{m^2}+4+\frac{\pi^2}{6}\right].
\end{align}
Then the one-loop anomalous dimension associated with this jet function is given by
\begin{align}
    \Gamma_J &= \frac{\alpha_e}{4\pi} \left( 4 \ln \frac{\mu^2}{m^2} + 2\right). 
\end{align}
It should be noted that the typical scale of the jet function is $\mu_j=m$, and that as matching coefficients of low energy effective theory, both hard and jet functions do not depend on the small transverse momentum.

The second line of Eq. \eqref{eq:fac_qT} represents the factorization of the massive soft function in Ref. \cite{Shao:2022stc}, which accounts for the contribution of real photon emissions. In the small $m$ limit, the massless soft function $S$ is defined in terms of soft Wilson lines
\begin{align}\label{eq:soft_wilson_line}
	S_{n_i}(x)=  \exp \left[-i e \int_{-\infty}^0 d s n_i \cdot A\left(x+s n_i\right)\right],
\end{align}
which describe a point-like source traveling along the path $x^\mu+s n_i^\mu$ with the light-like vector $n_i^2=0$. In the $r_\perp$-space we have the soft function
\begin{align}
	\tilde{S}(r_\perp,\Delta y) = \langle 0 |  \mathrm{\bar T} \left[ S^\dagger_{n_1}(r_\perp)S_{n_2}(r_\perp)\right] \mathrm{T} \left[ S^\dagger_{n_2}(0)S_{n_1}(0)\right] |0\rangle, 
\end{align}
where $n_{1,2}$ denote the directions of finial-state leptons. Expanding the Wilson line in the coupling perturbatively, the one-loop soft function is obtained as
\begin{align}
	\tilde{S}^{\rm NLO}(r_\perp,\Delta y) = 1 + e_0^2\int\frac{d^dk}{(2\pi)^{d-1}}\delta(k^2)\theta(k^0) \frac{2n_1\cdot n_2}{n_1\cdot k \, k \cdot n_2}e^{i k_\perp \cdot r_\perp},	
\end{align}
where $e_0$ is the bare electric charge, and $k$ is the momentum of the final-state photon. Note that $k_\perp$ is the photon transverse momentum with the beam directions which is different from the direction of $n_i$. After performing the momentum integral, we obtain
\begin{align}
	\tilde{S}^{\rm NLO}(r_\perp,\Delta y) = 1 + \frac{\alpha_e}{4\pi}\left[ \frac{4}{\epsilon^2} + \frac{4}{\epsilon} \ln \frac{\mu^2 r_\perp^2}{b_0^2 A_r}  + 2 \ln^2 \frac{\mu^2 r_\perp^2}{b_0^2A_r} + \pi^2 - 4 \ln A_r \ln(1-A_r) - 4\, \mathrm{Li}_2(A_r)  \right], 
\end{align}
with $A_r=M^2/(4P_\perp^2\cos^2\phi_r)$. In the $r_\perp$-space $\mu_r$ is chosen as the soft scale, and the anomalous dimension is 
\begin{align}\label{eq:adim_s_qT}
    \Gamma_S &= \frac{\alpha_e}{4\pi}\left( 8 \ln \frac{\mu^2r_\perp^2}{b_0^2} + 8 \ln \cos^2\phi_r - 8 \ln\frac{1+\cosh\Delta y}{2}\right). 
\end{align}
It is apparent that as $\phi_r$ approaches $\pi/2$ or $3\pi/2$, i.e., when the direction of $r_\perp$ becomes perpendicular to the lepton direction, the expression becomes divergent due to the presence of $\ln \cos^2\phi_r$. This divergence is connected to the rapidity divergence that dimensional regulators cannot regulate. We will elaborate on this further in the next subsection when we introduce the factorization formula for the acoplanarity distribution.

\begin{figure}[t]\centering
    \includegraphics[scale=0.45]{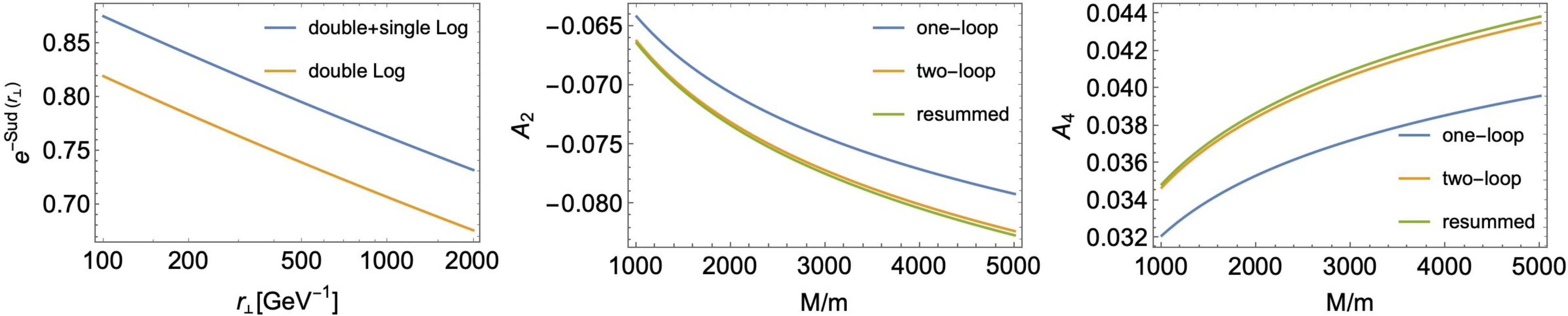}
    \caption{Sudakov factor Sud$(r_\perp)$ and azimuthal asymmetry $A_{2,4}$ in $q_\perp$ resummation formula (color online). Left panel: Sudakov factor for double logarithmic (yellow line) and double $+$ single logarithmic (blue line) contributions. The values of $M$, $m$, and $\Delta y$ are chosen to be $500$ MeV, $0.5$ MeV, and $0$, respectively. Middle panel: Azimuthal asymmetry $A_2$ shown for leading order (blue), next-to-leading order (yellow), and all-order (green) results. Right panel: Azimuthal asymmetry $A_4$ shown in the same colors as $A_2$.  }
    \label{fig:qTSud}
\end{figure}

The soft function describes large-angle long wave photons contribution, while the collinear-soft function $C_i$ captures contribution from the soft photon radiating close to the lepton direction, which is defined as
\begin{align}
	\tilde{C}_i(r_\perp,P_\perp,y_i,m) = \langle 0 |  \mathrm{\bar T} [ S^\dagger_{v_i}(r_\perp)S_{\bar n_i}(r_\perp)] \mathrm{T} [ S^\dagger_{\bar n_i}(0)S_{v_i}(0)] |0\rangle, 
\end{align}
where the soft Wilson line $S_{v_i}$ is defined in analogy with $S_{n_i}$ in Eq.~\eqref{eq:soft_wilson_line}, but with the light-like vector $n_i$ replaced with the time-like vector $v_i$, which is 
\begin{align}
	v_i^\mu=\frac{\omega_i}{m} \frac{n_i^\mu}{2}+\frac{m}{\omega_i} \frac{\bar{n}_i^\mu}{2},~~~\mathrm{with}~\omega_i = 2P_\perp\cosh y_i. 
\end{align}
At one loop, the perturbative expansion of collinear-soft function gives us
\begin{align}
	\tilde{C}_i^{\rm NLO}(r_\perp,P_\perp,y_i,m) = 1 + e_0^2\int\frac{d^dk}{(2\pi)^{d-1}}\delta(k^2)\theta(k^0) \left(\frac{2 v_i\cdot \bar n_i}{v_i\cdot k \, k \cdot \bar n_i} - \frac{ v_i \cdot v_i}{v_i\cdot k \, k \cdot v_i}\right)e^{i \bar n_i \cdot k \, n_i \cdot r_\perp/2},
\end{align}
then we obtain
\begin{align}
	\tilde{C}_i^{\rm NLO}(r_\perp,P_\perp,y_i,m) = 1 + \frac{\alpha_e}{4\pi}\left[ - \frac{2}{\epsilon^2} + \frac{2}{\epsilon}\left( 1 -  2\ln \mu R \right) - 4 \ln^2 \mu R + 4\ln \mu R - \frac{5\pi^2}{6} \right],
\end{align}
where $R = - i P_\perp  e^{\gamma_E} n_i \cdot r_\perp /(m \, r_\perp)$, and the anomalous dimension is 
\begin{align}\label{eq:adim_cs_qT}
    \Gamma_{C_{1,2}} &= \frac{\alpha_e}{4\pi}\left( - 4 \ln \frac{4P_\perp^2\mu^2r_\perp^2}{b_0^2 m^2} + 4 - 4 \ln \cos^2\phi_r \pm 4 i \pi  \right).  
\end{align}
We set the collinear soft scale as $\mu_c=\mu_r m/(2P_\perp)$. With the anomalous dimensions presented for all the ingredients, we now show that our factorized formula satisfies the consistency relations for the RG evolutions. The consistency equation reads
\begin{align}
	\Gamma_H + \Gamma_S + 2\Gamma_J	+\Gamma_{C_1} + \Gamma_{C_2} =0.
\end{align}
Based on the above discussions on the intrinsic scale and RG methods in SCET, we can obtain the expression for the all-order resummed cross section, and the Sudakov factor is given by
\begin{align}
	\mathrm{Sud}(r_\perp) = \int_{\mu_r}^M \frac{d\mu}{\mu} \Gamma_H + 2\int_{\mu_r}^m \frac{d\mu}{\mu} \Gamma_J + \int_{\mu_r}^{\mu_r m/(2P_\perp)} \frac{d\mu}{\mu} \Gamma_{C_{1}} + \int_{\mu_r}^{\mu_r m/(2P_\perp)} \frac{d\mu}{\mu} \Gamma_{C_{2}}, 
 \label{Res_all}
\end{align}
where we evolve hard, jet and collinear-soft functions from their intrinsic scale to $\mu_r$. After taking $\Delta y=0$, and neglecting the contribution from single logarithmic terms, we find
\begin{align}\label{eq:Sud_all}
	\mathrm{Sud}(r_\perp)\bigg|_{\mathrm{DL}, \Delta y=0} = \frac{\alpha_e}{\pi}\ln \frac{M^2}{m^2}\ln\frac{P_\perp^2}{\mu_r^2} + \frac{\alpha_e}{\pi}\ln \frac{M^2}{m^2}\ln4\cos^2\phi_r, 
\end{align}
where the first term on the right is consistent with the Sudakov factor given in Eq.~\eqref{eq:sud_DL} \cite{Hatta:2020bgy,Hatta:2021jcd,Shao:2022stc}. In order to investigate the contributions from single logarithms, in the left panel of Fig.~\ref{fig:qTSud} we present the Sudakov factor for only double logarithmic terms and both double and single logarithmic terms. We find that the single logarithmic corrections reduce the Sudakov suppression.

Moreover, the azimuthal angle-dependent terms that are enhanced in the small mass limit are also resummed into an exponential form, and the azimuthal angle correlation coefficients are given by
\begin{align}
    A_2 &\equiv \int_0^{2\pi} d\phi_r \frac{\cos2\phi_r}{\pi} \exp\left(-\frac{\alpha_e}{\pi}\ln \frac{M^2}{m^2}\ln4\cos^2\phi_r\right), \\
    A_4 &\equiv \int_0^{2\pi} d\phi_r \frac{\cos4\phi_r}{\pi} \exp\left(-\frac{\alpha_e}{\pi}\ln \frac{M^2}{m^2}\ln4\cos^2\phi_r\right). 
\end{align}
As a consistent check, we expand the above expression at one loop, and find they reproduce the coefficient $c_2$ and $c_4$ in the limit $M \gg m$ \cite{Shao:2022stc} as follows
\begin{align}
	&A_2^{\rm LO} = - \frac{\alpha_e}{\pi}\ln \frac{M^2}{m^2} \int_0^{2\pi} d\phi_r \frac{\cos2\phi_r}{\pi} \ln 4 \cos^2\phi_r =  - \frac{2\alpha_e c_2 }{\pi},~~~~\mathrm{with}~ c_2 \approx \ln \frac{M^2}{m^2}, \\
	& A_4^{\rm LO} =  - \frac{\alpha_e}{\pi}\ln \frac{M^2}{m^2} \int_0^{2\pi} d\phi_r \frac{\cos4\phi_r}{\pi} \ln 4\cos^2\phi_r =   \frac{\alpha_e c_4 }{\pi},~~~~~~~~\mathrm{with}~ c_4 \approx \ln \frac{M^2}{m^2},
 \label{azdep}
\end{align}
where as the azimuthal correlation first appears at one loop, we refer to the corresponding coefficients as the leading-order (LO) coefficients. Besides, we can use the all-order formula \eqref{eq:Sud_all} to explore the azimuthal angular correlation coefficients at higher orders. The middle and right panels of Fig. \ref{fig:qTSud} show $A_{2,4}$ at LO (one-loop), NLO (two-loop), and all orders. It is evident that the high-order corrections enhance the azimuthal asymmetry, and the degree of enhancement depends on the scale hierarchy between $M$ and $m$. In the typical RHIC kinematic regions, $A_2$ and $A_4$ increase by about $5\%$ and $10\%$, respectively.

\subsection{Factorization formula at low $\alpha$}

In the previous subsection we derived a factorization formula for low values of $q_\perp$. Equations \eqref{eq:adim_s_qT} and \eqref{eq:adim_cs_qT} show that the anomalous dimensions become divergent as the direction of $r_\perp$ becomes perpendicular to the lepton direction. If we choose that the direction of lepton transverse momentum is along the $y$-axis and $r_\perp=r_x$, then the anomalous dimensions in Eqs. \eqref{eq:adim_s_qT} and \eqref{eq:adim_cs_qT} diverge in dimensional regularization. As a result, we need to re-derive a factorization formula in the small $\alpha$ limit since the acoplanarity $\alpha$ is reconstructed by $q_x$ (or $r_x$ in the conjugate Fourier space).

Since the hard and jet functions in Eq. \eqref{eq:fac_qT} are matching coefficients that are independent of the specific observable, they should be the same in the factorization formula for the $\alpha$ distribution. In other words, only the soft and collinear-soft functions need to be modified in this case. As $\alpha\to 0$, the factorization formula should be expressed as
\begin{align}
	\frac{d\sigma(\alpha)}{ d\mathcal{P.S.}} =& 2 P_\perp H(M,\mu) J^2(m,\mu) \int d l_x d k_{1,x} d k_{2,x}  \frac{d\sigma_0(q_x-l_x - k_{1x }- k_{2x} )}{ d\mathcal{P.S.}} \notag \\
 &\times S(l_x,\Delta y,\mu,\nu)  C_1 ( k_{1x },P_\perp,y_1,m,\mu,\nu ) C_2 ( k_{2x },P_\perp,y_2,m,\mu,\nu ),  
\end{align}
where only a one-dimensional Fourier transformation is needed, as explained in Section \ref{sec:DL}. The soft and collinear-soft functions exhibit different divergence structures from those in Eq. \eqref{eq:fac_qT}. Specifically, the naive separation of soft and collinear-soft momentum regions is not well-defined without additional regulators. The modified factorization formula for the $\alpha$ distribution takes into account these extra divergences. The variable $\nu$ denotes the scale introduced by the dimensionless rapidity regulator. In this study, we will utilize the analytical regulator introduced in Refs. \cite{Becher:2010tm,Becher:2011dz,Bell:2018oqa}, and alternative regulators can be found in Refs. \cite{Collins:2011zzd,Echevarria:2011epo,Echevarria:2012js,Chiu:2011qc,Chiu:2012ir,Li:2016axz,Ebert:2018gsn}. 
 
After performing the one-dimensional Fourier transformation, the operator definition of the soft function is given by
\begin{align}
	\tilde{S}(r_x,\Delta y) = \langle 0 |  \mathrm{\bar T} \left[ S^\dagger_{n_1}(r_x)S_{n_2}(r_x)\right] \mathrm{T} \left[ S^\dagger_{n_2}(0)S_{n_1}(0)\right] |0\rangle, 
\end{align}
where $r_x$ points along $x$-direction which is perpendicular to the direction of final state leptons in the $y$-$z$ plane. Therefore the NLO soft function is expressed as 
\begin{align}
	\tilde{S}^{\rm NLO}(r_x,\Delta y) = 1 + e_0^2\int\frac{d^dk}{(2\pi)^{d-1}}\delta(k^2)\theta(k^0) \left(\frac{\nu}{2k^0}\right)^\eta \frac{2n_1\cdot n_2}{n_1\cdot k \, k \cdot n_2}e^{i k_x  r_x},	
\end{align} 
where we introduce the rapidity regulator to regularize the rapidity divergence. In order to evaluate this integral, it is convenient to boost two light-like vectors $n_{1,2}$ into their center-of-mass frame. Since such a boost operation can be performed in the $y$-$z$ plane, the Fourier exponent function is not changed. As a result, only the rapidity regulator transforms as
\begin{align}
	\left(\frac{\nu}{2k^0}\right)^\eta \to \left(\frac{\nu}{2k^0} \sqrt{\frac{n_1\cdot n_2}{2}} \right)^\eta,
\end{align}
Therefore, we have the NLO soft function
\begin{align}
	\tilde{S}^{\rm NLO}(r_x,\Delta y) = 1 + \frac{\alpha_e}{4\pi}\left[ \frac{4}{\epsilon^2} - 4\left( \ln \frac{\mu^2r_x^2}{b_0^2}+ \frac{1}{\epsilon}\right)\left( \frac{2}{\eta} + \ln \frac{n_1\cdot n_2 \nu^2}{2\mu^2} \right) - 2 \ln^2 \frac{\mu^2r_x^2}{b_0^2} - \frac{\pi^2}{3}  \right],
\end{align}
where the $1/\eta$ poles comes from rapidity divergences. In order to resum large logarithms associated with rapidity divergences, one can apply collinear anomaly \cite{Becher:2010tm,Becher:2011dz} methods. 

Similarly to the soft function, the operator definition of the collinear soft function takes the form
\begin{align}
	\tilde{C}_i(r_{x},P_\perp,y_i,m) =  \langle 0 |  \mathrm{\bar T} [ S^\dagger_{v_i}(r_{\perp i})S_{\bar n_i}(r_{\perp i})] \mathrm{T} [ S^\dagger_{\bar n_i}(0)S_{v_i}(0)] |0\rangle, 
\end{align}
where $r_{\perp i}$ is perpendicular to the direction of the lepton. At one loop, we have 
\begin{align}
	\tilde{C}^{\rm NLO}_i(r_{\perp i},P_\perp,y_i,m) = 1 + e_0^2\int\frac{d^dk}{(2\pi)^{d-1}}\delta(k^2)\theta(k^0) \left(\frac{\nu}{\bar n_i\cdot k}\right)^\eta \left(\frac{2 v_i\cdot \bar n_i}{v_i\cdot k \, k \cdot \bar n_i} - \frac{ v_i \cdot v_i}{v_i\cdot k \, k \cdot v_i}\right)e^{i  k_{\perp i}  \cdot r_{\perp i} },
\end{align}
where the small component of the momentum $n_i\cdot k$ in the rapidity regulator is expanded out and only the large component $\bar n_i\cdot k$ is retained since $2k^0=n_i\cdot k+\bar n_i\cdot k$ in the light-cone coordinate. 
Therefore
\begin{align}
	\tilde{C}^{\rm NLO}_i(r_{\perp i},P_\perp,y_i,m) = 1 + \frac{\alpha_e}{4\pi} \left[ - \frac{2}{\epsilon^2} + \frac{2}{\epsilon} - 2\left( \ln \frac{\mu^2 r_x^2}{b_0^2} + \frac{1}{\epsilon}\right)\left( \ln\frac{\omega_i^2 \mu^2}{m^2 \nu^2}-\frac{2}{\eta} \right) + \ln^2\frac{\mu^2 r_x^2}{b_0^2} + 2\ln\frac{\mu^2 r_x^2}{b_0^2} + \frac{25\pi^2}{6}\right].
\end{align}
It is clear to see that the rapidity poles are canceled after combining soft and collinear-soft functions. Explicitly, we have 
\begin{align}
	\tilde{S} \tilde{C}_1 \tilde{C}_2 = 1 + \frac{\alpha_e}{4\pi} \left( - 4 \ln\frac{\mu^2r_x^2}{b_0^2} \ln \frac{M^2}{m^2} + 4 \ln\frac{\mu^2r_x^2}{b_0^2} + 8 \pi^2 \right) + \mathcal{O}(\alpha_e^2),
\end{align}
where the UV poles have been removed by the $\overline{\rm MS}$ subtraction scheme. Besides, the logarithm of the ratio between $M$ and $m$ can not be resummed by standard RG equations, and this problem is referred to as the collinear anomaly, where the extra large logarithms are resummed by the colliner anomaly factor. Explicitly, we define 
\begin{align}\label{eq:refac_ca}
\tilde{S} \tilde{C}_1 \tilde{C}_2 = \left(\frac{M^2}{m^2}\right)^{-F(r_x,\mu)} W(r_x,\mu),
\end{align}
where the anomaly exponent $F$ depends only on $r_x$ and the renormalizaton scale $\mu$, and its one-loop expression is
\begin{align}
    F(r_x,\mu) = \frac{\alpha_e}{\pi} \ln \frac{\mu^2r_x^2}{b_0^2} + \mathcal{O}(\alpha_e^2).
\end{align}
It satisfies the following RG equations
\begin{align}
    \frac{d}{d\ln\mu}F(r_x,\mu)=\frac{2\alpha_e}{\pi}.
\end{align}
In Eq.~\eqref{eq:refac_ca} we have introduced the remainder function $W(r_x,\mu)$ which also depends only on $r_x$ and $\mu$
\begin{align}
    W^{\rm NLO}(r_x,\mu) = 1 + \frac{\alpha_e}{4\pi} \left(  4 \ln\frac{\mu^2r_x^2}{b_0^2} + 8 \pi^2 \right),
\end{align}
with the one-loop anomalous dimension as $\Gamma_W=2\alpha_e/\pi$. The typical scale in $F$ and $W$ functions is $\mu_{rx}$, and the RG consistency can be easily verified at one loop. After evolving the hard and jet functions to $\mu_{rx}$, we obtain the all-order resummation formula for $\alpha$ distribution, where the Sudakov factor is expressed as
\begin{align}
	\mathrm{Sud}_a(r_x) & = \int_{\mu_{rx}}^M \frac{d\mu}{\mu} \Gamma_H + 2\int_{\mu_{rx}}^m \frac{d\mu}{\mu} \Gamma_J \theta(m-\mu_{rx}), \label{acop} \\ \notag 
	& = \frac{\alpha_e}{2\pi} \left[\left( \ln^2 \frac{M^2}{\mu_{rx}^2} - 3 \ln \frac{M^2}{\mu_{rx}^2} \right) - \left( \ln^2 \frac{m^2}{\mu_{rx}^2} - \ln \frac{m^2}{\mu_{rx}^2}\right) \theta(m-\mu_{rx})\right],
\end{align}
where the double logarithmic terms are consistent with the expression in Eq. \eqref{eq:suda_DL}.  

\begin{figure}[t]\centering
    \includegraphics[scale=0.45]{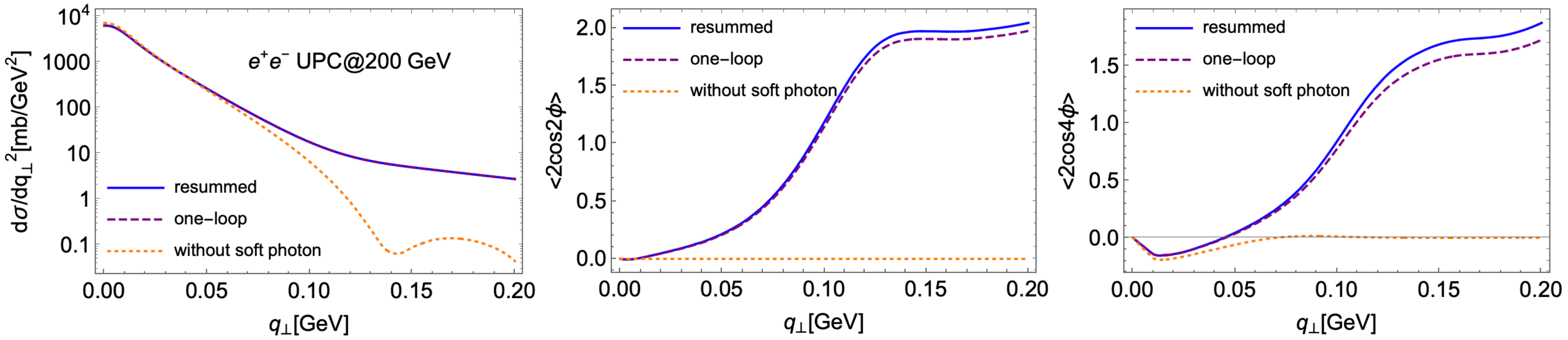}
    \caption{Di-electron production in unrestricted UPCs in Au+Au collisions at  RHIC energy. The following kinematic cuts are imposed: the electrons' rapidities $|y_{1,2}|<1$, transverse momentum $P_\perp >200$ MeV, and the invariant mass of the electron  pair $450~\text{MeV}< M < 760~\text{MeV}$. The blue solid lines stand for the fully resummed results from Eq.\eqref{Res_all}, and the purple dashed lines represent the results with the azimuthal dependent part being treated at the one loop order. The results without soft photon radiation effect are shown with the dotted orange lines. Left panel: azimuthal averaged differential cross sections; middle panel: $\langle\cos(2\phi)\rangle$ azimuthal asymmetry; right panel: $\langle\cos(4\phi)\rangle$ azimuthal asymmetry.}
    \label{fig2}
\end{figure}

\begin{figure}[t]\centering
    \includegraphics[scale=0.45]{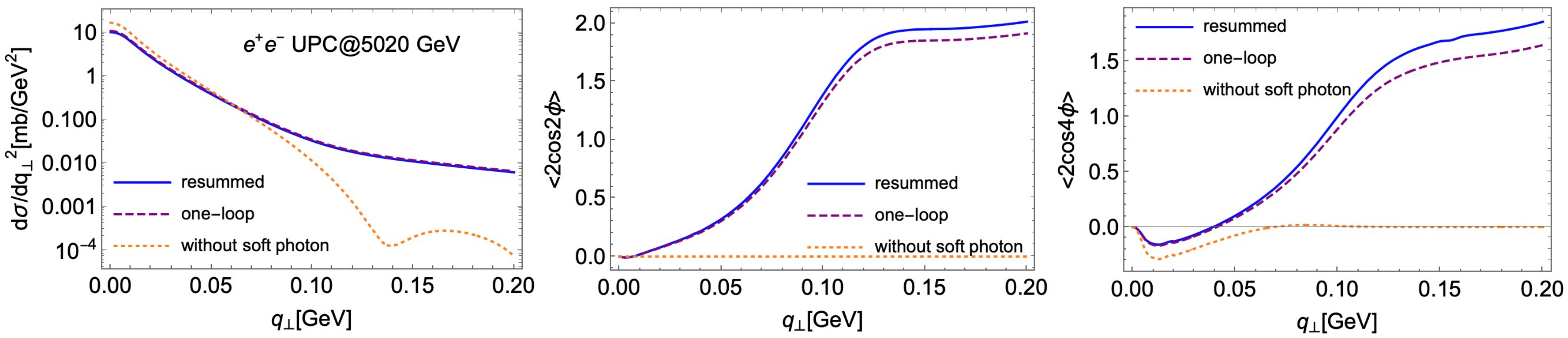}
    \caption{Di-electron production in unrestricted UPCs in Pb+Pb collisions at  LHC energy. The following kinematic cuts are imposed: the electrons' rapidities $|y_{1,2}|<0.8$ and the invariant mass of the di-electron $10~\text{GeV}< M < 20~\text{GeV}$. The blue solid lines stand for the fully resummed results from Eq.\eqref{Res_all}, and the purple dashed lines represent the results with the azimuthal dependent part being treated at the one loop order. The results without soft photon radiation effect is shown with the dotted orange lines. Left panel: azimuthal averaged differential cross sections; middle panel: $\langle\cos(2\phi)\rangle$ azimuthal asymmetry; right panel: $\langle\cos(4\phi)\rangle$ azimuthal asymmetry.}
    \label{fig3}
\end{figure}

\section{numerical results}
We now discuss  the model input  used in the numerical evaluations.
 It is convenient to perform the numerical calculation with the electromagnetic form factor  taken from the STARlight MC generator~\cite{Klein:2016yzr},
\begin{eqnarray}
F(|\vec k|)=\frac{4\pi \rho^0}{|\vec k|^3 A}\left [ \sin(|\vec k|R_A)-|\vec k|R_A \cos(|\vec k|R_A)\right ]\frac{1}{a^2 \vec k^2+1},
\label{ff}
\end{eqnarray}
where   $a=0.7$ fm, and $\rho^0$ is a normalization factor. The nucleus radius is chosen to be $R_A=1.1 A^{1/3}$fm for Au and Pb targets.
This parametrization is numerically very close to the Woods-Saxon distribution.

 The  azimuthal asymmetries, i.e., the average value of  $\cos 2n \phi$ are defined as,
\begin{eqnarray}
\langle \cos(2n\phi) \rangle &=&\frac{ \int \frac{d \sigma}{d {\cal P.S.}} \cos (2n \phi) \ d {\cal P.S.} }
{\int \frac{d \sigma}{d {\cal P.S.}}  d {\cal P.S.}}.
\end{eqnarray}
We compute both the azimuthal independent cross sections and the asymmetries for the unrestricted UPC case where we simply integrated the impact parameter over the range $[2R_{\rm WS}, \infty)$, with the nucleus radius $R_{\rm WS}$ being 6.4 fm for Au and 6.68 fm for Pb. 

\begin{figure}[t]\centering
    \includegraphics[scale=0.55]{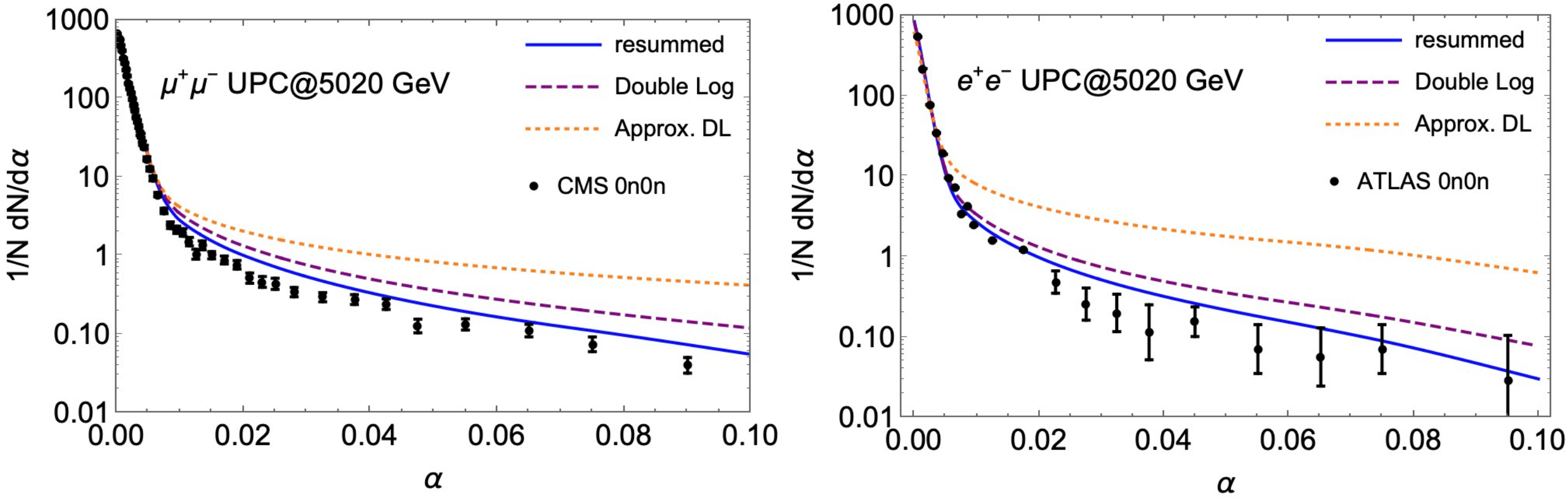}
    \caption{The normalized cross sections of di-lepton production are plotted as the function of $\alpha$ (color online). Left panel: di-muon production in Pb+Pb collisions for the 0n0n case, with the kinematic cutoff: leptons' rapidities $|y_{1,2}|<2.4$, transverse momentum $P_\perp >3.5$ GeV, and the invariant mass of the di-muon  $8~\text{GeV}< M < 60~\text{GeV}$. The CMS data displayed in the figure is taken from \cite{CMS:2020skx}.
    Right panel: di-electron production  in Pb+Pb collisions for the 0n0n case, with the kinematic cutoff: leptons' rapidities $|y_{1,2}|<0.8$ and the invariant mass of the di-electron  $10~\text{GeV}< M < 20~\text{GeV}$. The ATLAS data shown in the figure is taken from \cite{ATLAS:2022srr}.
    The blue solid lines stand for the fully resummed results from Eq.\eqref{acop}, the purple dashed lines represent the leading double logarithm resummed results obtained  using Eq.\eqref{eq:suda_DL}. The acoplanarity distribution reconstructed from the resummed $q_\perp$ distribution given by Eq.~\eqref{res1} and  Eq.~\eqref{eq:sud_DL} is shown with the dotted orange lines. }
    \label{fig4}
\end{figure}

The azimuthal independent and dependent cross sections are  plotted as the function of $q_\perp$ 
 at RHIC energy in Fig. \ref{fig2} and LHC energy in Fig.~\ref{fig3}. It is clear to see that at relatively high $q_\perp$, the perturbative tail generated by soft photon radiation dominates over the lepton pair transverse momentum spectrum determined by the coherent photon primordial $k_\perp$ distribution. In this work,  both the azimuthal independent and dependent leading logarithms are resummed into an exponential form, whereas in the previous work~\cite{Hatta:2020bgy,Hatta:2021jcd}, we only resummed azimuthal independent logarithm to all orders and treat the azimuthal dependent piece at the fixed order.  We numerically compare the results computed from these two resummation scheme. The difference between these two  methods becomes manifest  when evaluating the azimuthal asymmetries in the large $q_\perp$ region, in particular for $\cos 4\phi$ azimuthal asymmetry. It would be interesting to test such  resummation effect in the future experiment. 

 The acoplanarity distributions computed at LHC energy for both di-electron and di-muon production are displayed in Fig.~\ref{fig4}.  To avoid the possible contribution from incoherent photons which could play a role in the large $\alpha$ region, we only make numerical estimations for the 0n0n events in which no neutron is emitted after the EM interaction occurs. 
For the 0n0n event, the impact parameter dependence of the cross section is weighted with an $b_\perp$ distribution (see the review article~\cite{Miller:2007ri} and references therein),
 \begin{eqnarray}
2 \pi \int_{2R_{WS}}^{\infty} b_\perp db_\perp P^2(b_\perp) d \sigma(b_\perp, \ ...),
\end{eqnarray}
 where the probability $P(b_\perp)$ for the 0n event is commonly parameterized as~\cite{Baur:1998ay},
\begin{eqnarray}
P(b_\perp)= \exp \left [-5.45*10^{-5}\frac{Z^3(A-Z) }{A^{2/3} b_\perp^2} \right ].
\end{eqnarray}
The theoretical calculation is consistent with the both ATLAS and CMS low $\alpha$ data. However, in the relatively large $\alpha$ region,  our numerical results clearly overshoot the experimental data. The inclusion of the leading single logarithm contribution in the resummation formalism does relieve the tension between the experimental data and the theory calculation to some extent.  The possible origin of this discrepancy is that the collinear physics is not fully captured in our resummation formalism. We will address this point in the future work. In the meantime, we also reconstruct the  acoplanarity using the resummed $q_\perp$ distribution given in Eq.~\eqref{res1}.  
 The measured $\alpha$ distribution is obviously not in favor of this approach as shown in Fig.~\ref{fig4}.

\section{conclusion}
We study the azimuthal angular correlations of high-$q_\perp$ lepton pairs produced in UPCs, which are mainly generated by soft photon radiation in the final state. We show that the resummation of soft photon radiation has different formulations for the lepton pair $q_\perp$ distribution and the acoplanarity distribution, and that it is not valid to infer the acoplanarity distribution from the resummed $q_\perp$ distribution. Within the SCET framework, we perform the all order  resummation for both observables up to the single leading logarithm accuracy.
Our results show that the $q_\perp$ dependent azimuthal asymmetries are not very sensitive to sub-leading resummation effects, but the leading single logarithm contribution is essential to describe the acoplanarity data from ATLAS and CMS. However, our calculations still exceed the data for large $\alpha$. This discrepancy certainly warrants further investigation.  Nevertheless, we conclude that the process of lepton pair production in UPCs provide a great opportunity to test the resummation formalism  through angular correlations, thanks to the high coherent photon luminosity and the high angular resolution of modern detectors~\cite{Gao:2022bzi}. The resummation formalism presented here can be extended to  study the angular correlations in  the   diffractive productions of di-jet, jet-hadron and hadron-hadron  in UPCs. We leave these for future studies.

\section*{Acknowledgments}
We thank Chi Yang, Shuai Yang, Xiao-feng Wang and Tianbo Liu for helpful discussions. 
D.Y.S.~is supported by the National Science Foundations of China under Grant No.~12275052 and No.~12147101 and the Shanghai Natural Science Foundation under Grant No. 21ZR1406100. J. Zhou has been supported by the National  Science Foundations of China under Grant No.\ 12175118. Y. Zhou has been supported by the Natural  Science Foundation of Shandong Province under Grant No. ZR2020MA098. C. Zhang has been supported by the National Science Foundations of China under Grant No.\ 12147125.

\appendix

\section{A Heuristic  derivation  of the  Sudakov factor $ \mathrm{Sud}(r_\perp)$ }
The sub-leading logarithm contribution to the Sudakov factor  $\mathrm{Sud}(r_\perp)$ can be derived in an alternative way. We start with discussing the azimuthal angular dependent part. 
The soft factor in the leading logarithm approximation can be expanded as~\cite{Hatta:2021jcd},
\begin{eqnarray}
S(l_\perp)=\frac{\alpha_e}{\pi^2} \frac{1}{l_\perp^2} \ln \frac{M^2}{m^2} \left \{ 1+2\cos 2\phi +2\cos4\phi +2\cos 6\phi+... \right \},
\end{eqnarray}
where $\phi$ is the azimuthal angle between soft photon transverse momentum $l_\perp$ and $P_\perp$.
 Our task is to Fourier transform the soft factor to $r_\perp$ space 
$S(r_\perp)=\int d^2l_\perp  e^{ir_\perp \cdot l_\perp} S(l_\perp) 
$.
 With the help of the Jacobi-Anger expansion,
\begin{eqnarray} 
e^{iz\cos(\phi)}=J_0(z) +2 \sum_{n=1}^{\infty} i^n J_n(z) \cos(n\phi)\,,
\end{eqnarray}
and the integration formula,
\begin{eqnarray} 
\int_0^\infty \frac{d |q'_\perp|}{|q'_\perp|} J_n(|q'_\perp||b_\perp|)=\frac{1}{n} \ , \label{jn}
\end{eqnarray}
 one arrives at,
\begin{eqnarray} 
S(r_\perp)=\frac{\alpha_e}{\pi} \ln \frac{M^2}{m^2}  4\sum_{n=1}^{\infty} \frac{i^{2n} }{2n} \cos(2n\phi_r)=-\frac{\alpha_e}{\pi}\ln \frac{M^2}{m^2} \ln \left[ 2+2\cos (2\phi_r) \right ] \ .
\end{eqnarray}
where $\phi_r$ is the azimuthal angle between $r_\perp$ and $P_\perp$.
This result is consistent with the second term in Eq.~\eqref{eq:Sud_all}.

Now we turn to discuss the derivation of the single logarithm terms.
For simplicity we consider a special case $\Delta y=0$. The part of the single logarithm contribution purely comes from the virtual correction. 
The leading logarithm virtual correction can be expressed as (see for example~\cite{Liu:2020rvc,Liu:2021jfp}),
\begin{eqnarray} 
\frac{\alpha_e}{2\pi^2} \int_0^M
\frac{d^2 l_\perp}{l_\perp^2+\frac{(1-z)^2}{z^2} m^2}\int dz  \frac{1+z^2}{1-z},
\end{eqnarray}
in the frame where the electron momentum is chosen to be the light-cone direction. $z$ stands for the longitudinal momentum fraction of electron carried by virtual photon. The UV cutoff is chosen to be the lepton pair invariant mass. The virtual contribution from the soft region has already been combined with the real correction to form the leading double logarithm contribution. Therefore, we have to subtract the soft region contribution, 
\begin{eqnarray} 
\frac{\alpha_e}{2\pi^2} \int_0^M
\frac{d^2 l_\perp}{l_\perp^2+\frac{(1-z)^2}{z^2}m^2}\int dz  \left [ \frac{1+z^2}{1-z}-  \frac{2}{1-z} \right ]=\frac{3}{4} \frac{\alpha_e}{\pi}\ln\frac{M^2}{m^2} +\mathcal{O}\left (\frac{1}{\ln\frac{M^2}{m^2}} \right ). \label{single1}
\end{eqnarray}

Another contribution to the single logarithm term is from the diagram where soft photon connects two electron lines or two  positron lines. After applying the Eikonal approximation, one has,
\begin{eqnarray} 
\int \frac {d^3q}{ 2q^0}  \frac{m^2}{(q \cdot P)^2}=\int \frac{d q_\perp^2}{4q_\perp^2 P_\perp^2}\int dy d \phi \frac{m^2}{\left [ \sqrt{1+\frac{m^2}{P_\perp^2}}\cosh(y)-\cos \phi \right ]^2}=\int \frac{d q_\perp^2}{4q_\perp^2 P_\perp^2}\int dy  \frac{m^2  2 \pi \sqrt{1+\frac{m^2}{P_\perp^2}}\cosh(y) }{\left [ \left (1+\frac{m^2}{P_\perp^2}\right )\cosh^2(y)-1 \right ]^{\frac{3}{2}}}.
\end{eqnarray}
After changing the variable $e^y=z$, the above $y$ integration can be readily carried out,
\begin{eqnarray} 
e^2 \int \frac {d^3q}{ (2\pi)^3 2q^0}  \frac{m^2}{(q \cdot P)^2}\approx  \frac{\alpha_e}{2\pi^2} \int \frac{d^2 q_\perp}{q_\perp^2},
\end{eqnarray}
where  the terms suppressed by the power of $m^2/P_\perp^2$ have been neglected. Now we combine the virtual and real corrections together,
\begin{eqnarray} 
\frac{\alpha_e}{2\pi^2}\int^M \frac{d^2 q_\perp}{q_\perp^2}\left (1-e^{iq_\perp \cdot r_\perp} \right )\approx \frac{\alpha_e}{2\pi} \ln \frac{M^2}{\mu_r^2}. \label{single2}
\end{eqnarray}
The sum of Eq.~\eqref{single1} and Eq.~\eqref{single2} gives the full single logarithm contribution from each lepton line to the Sudakov factor $ \mathrm{Sud}(r_\perp)$.

\section{A Heuristic  derivation  of the  Sudakov factor $ \mathrm{Sud_a}(r_x)$ }
The Sudakov factor $ \mathrm{Sud_a}(r_x)$ can be reproduced by isolating the large logarithm contributions from the collinear splitting function for electron. In the collinear limit, the electron fragmentation function  at the leading order reads (see for example~\cite{Liu:2020rvc,Liu:2021jfp}),
\begin{eqnarray} 
\frac{\alpha_e}{2\pi^2} \int_0^M
\frac{d^2 l_\perp}{l_\perp^2+\frac{(1-z)^2}{z^2} m^2}\int dz  \frac{1+z^2}{1-z}+  \mathrm{virtual \ correction},
\label{dglap}
\end{eqnarray}
where $l_\perp$ is perpendicular to the outgoing electron momentum $p_1$. The X-component of $l_\perp$, i.e., $l_x$, is chosen such that it satisfies the following conditions,
\begin{eqnarray} 
p_2 \cdot l_x=0, \  \  \ p_{2\perp} \cdot l_x=0,   \ \ \ p \cdot l_x=0, \  \ \ n \cdot l_x=0,
\end{eqnarray}
where $p^\mu$ and $n^\mu$ are the commonly defined light-cone vectors along the beam direction in the lab frame.
 The acoplanarity is determeind by the ratio of $l_x$ and $p_{2\perp}=p_{2\perp,y}$ where $p_2$ is positron momentum. Since $l_y$ goes unobserved, it needs to be integrated out in the end of the calculations. Note that $l_y$ is generally not perpendicular to the light-cone momenta $p^\mu$ and $n^\mu$.

The light-cone divergence $z \rightarrow 1$ in Eq.~\eqref{dglap} can be cured by taking into account the exact kinematics.
According to the on-shell condition of radiated photon, one has,
\begin{eqnarray} 
\frac{l_\perp^2}{2\bar P ^-}<l^+<P^+,
\end{eqnarray}
where $P^+$ and $\bar P^-$ stand for the light-cone vectors along the lepton momentum direction instead of these along the beam direction. We take the collinear photon emission along $P^+$ direction as an example. 
To avoid double counting, we further require 
\begin{eqnarray} 
l^-<l^+,
\end{eqnarray}
which leads to the constraint,
\begin{eqnarray} 
\frac{\sqrt{2}|l_\perp|}{2}<l^+<P^+.
\end{eqnarray}
This converts to the integration limits for $z$ which is  specified as,
\begin{eqnarray} 
&&\frac{\alpha_e}{2\pi^2} \int_0^M
\frac{d^2 l_\perp}{l_\perp^2+\frac{(1-z)^2}{z^2} m^2}\int_0^{1-\frac{\sqrt{2}}{2} \frac{l_\perp}{P^+}} dz  \frac{1+z^2}{1-z}\nonumber \\&&
\approx \frac{\alpha_e}{2\pi^2} \int_0^M d^2 l_\perp \left [\int_0^{1} dz
\frac{1}{l_\perp^2+\frac{(1-z)^2}{z^2} m^2}   \frac{1+z^2}{(1-z)_+}+\int_0^{1- \sqrt{\frac{l_\perp^2}{M^2}} }dz \frac{1}{l_\perp^2+\frac{(1-z)^2}{z^2} m^2} \frac{2}{1-z} \right ] \nonumber \\ &&
\approx \frac{\alpha_e}{2\pi^2} \int_0^{1} dz \int_0^M
\frac{d^2 l_\perp}{l_\perp^2}   \frac{1+z^2}{(1-z)_+}+\frac{\alpha_e}{2\pi^2} \int_0^M
\frac{d^2 l_\perp}{l_\perp^2}\ln \frac{m^2+M^2}{l_\perp^2+m^2}.
\end{eqnarray}
The last term in the third line gives rises to the leading double logarithm contribution. Combining with the virtual correction, one has,
\begin{eqnarray} 
\int_0^M
\frac{d^2 l_\perp}{l_\perp^2}\ln \frac{M^2}{l_\perp^2+m^2}  \left [  e^{il_\perp \cdot r_\perp }-1\right]=\int_0^M
\frac{d^2 l_\perp}{l_\perp^2}\left \{\ln \frac{M^2}{l_\perp^2} +\ln \frac{l_\perp^2}{l_\perp^2+m^2}  \right \}\left [  e^{il_\perp \cdot r_\perp }-1\right].
\end{eqnarray}
As explained earlier, we should only make 
 the Fourier transform with respect to $l_x$, and integrated out $l_y$. The integration over $l_y$ can be easily achieved by setting $r_y=0$ at the end of the calculations. 
It is straightforward to carry out $l_\perp$ integration,
\begin{eqnarray} 
\int_0^M
\frac{d^2 l_\perp}{l_\perp^2}\ln \frac{M^2}{l_\perp^2} \left [  e^{il_\perp \cdot r_\perp }-1\right]\approx -\frac{\pi}{2} \ln^2 \frac{M^2}{\mu_r^2},
\end{eqnarray}
and 
\begin{eqnarray} 
\int_0^M
\frac{d^2 l_\perp}{l_\perp^2}\ln \frac{\l_\perp^2}{l_\perp^2+m^2}  \left [  e^{il_\perp \cdot r_\perp }-1\right] \approx  -\frac{\pi}{2}  \ln^2 \frac{m^2}{\mu_r^2} \theta(m-\mu_r),
\end{eqnarray}
where we only keep the leading logarithm contributions. After carrying out $l_y$ integration, $r_y$ is fixed to be 0. Correspondingly, $\mu_r$ is converted into $\mu_x$. These two double logarithm terms  can be promoted to an exponential form after carrying out all order resummation. 

Now we consider the collinear part that is free of the light-cone divergence,
\begin{eqnarray} 
&&\frac{\alpha_e}{2\pi^2} 
\frac{1}{l_\perp^2}  \frac{1+z^2}{(1-z)_+}-\frac{\alpha_e}{2\pi^2} \delta^2(l_\perp) \delta(1-z) \int
\frac{d^2 k_\perp}{k_\perp^2} \int_0^{1} d\xi  \frac{1+\xi^2}{(1-\xi)_+}  \nonumber \\ &&=
\frac{\alpha_e}{2\pi^2} 
\frac{1}{l_\perp^2} \left [ \frac{1+z^2}{(1-z)_+}+\frac{3}{2} \delta(1-z) \right ] -\frac{\alpha_e}{2\pi^2}\frac{3}{2} \delta(1-z) \left [ \frac{1}{l_\perp^2} -\delta^2(l_\perp)  \int
\frac{d^2 k_\perp}{k_\perp^2}  \right ],
\label{last1}
\end{eqnarray}
where it is safe to neglect $\frac{(1-z)^2}{z^2} m^2$ in the denominator as the integration is no longer dominated by the region $z\rightarrow  1$. The last two terms proportional to $\delta(1-z)$ can be resummed into an exponential form after  making the Fourier transform. In $r_\perp$ space, it reads,
\begin{eqnarray} 
-\delta(1-z)   \frac{\alpha_e}{2\pi^2}\frac{3}{2} \int_0^M 
\frac{d^2 l_\perp}{l_\perp^2} \left ( e^{ir_\perp \cdot l_\perp}-1 \right )=\delta(1-z) \frac{\alpha_e}{2\pi}\frac{3}{2}\ln \frac{M^2}{\mu_r^2},
\end{eqnarray}
which contributes to the single leading logarithm term in the Sudakov factor $ \mathrm{Sud_a}(r_x)$.  The term involving  the DGLAP splitting kernel in Eq.~\eqref{last1} should  be absorbed into the renormalized electron Fragmentation-PDF. Another single logarithm term $\frac{\alpha_e}{2\pi} \ln \frac{m^2}{\mu_r^2} \theta(m-\mu_r)$ receives the contribution from the diagrams with the soft photon connecting two electron lines or two positron lines in the cutting graphs. The derivation of this term is rather straightforward.  We thus reproduce both the double and the single logarithm terms in the Sudakov factor $\mathrm{Sud_a}(r_x)$ given in Eq.~\eqref{acop}.

\bibliography{ref}

\end{document}